\documentclass[12pt]{article}
\usepackage{amsmath}

\input{tcilatex}

\begin{document}

\begin{center}
\textbf{NEW STOCHASTIC CALCULUS}

\bigskip

Moawia Alghalith

malghalith@gmail.com

\bigskip
\end{center}

ABSTRACT. We present new stochastic differential equations, that are more
general and simpler than the existing Ito-based stochastic differential
equations. As an example, we apply our approach to the investment
(portfolio) model.

\bigskip

Key words: stochastic calculus, Ito calculus, stochastic PDEs, stochastic
process, portfolio, investment.

\pagebreak

\textbf{1. Introduction}

The literature on stochastic processes relied mainly on Ito's rule to derive
stochastic differential equations such as the Hamilton-Jacobi partial
differential equation. Examples include Siu (2012), Bayraktar and Young
(2010), Alghalith (2009), and Focardi and Fabozzi (2004), among many others.

It is well-known that Ito's rule is based on the assumptions of normality
and Levy processes such as Wiener processes (Brownian motions). In addition,
Ito's rule assumes convergence in probability. Moreover, the coefficients in
Ito's formula are not necessarily constant, which makes the formula less
convenient.

In this paper, we provide general and simple stochastic differential
equations that overcome the limitations of the existing differential
equations. In doing so, we derive stochastic differential equations that,
first, do not assume normality or any particular probability distribution.
Secondly, we do not assume a Levy process or any particular process.
Thirdly, we do not assume convergence. Moreover, these stochastic
differential equations have constant coefficients. This provides analytical
convenience.

\textbf{2. The model}

We consider the stochastic process $f\left( t,\xi \left( t\right) \right) ,$
where $\xi $ is stochastic and $t$ denotes time. Without loss of generality,
we define $\xi \left( t\right) =\bar{\xi}\left( t\right) +\sigma \left(
t\right) \eta \left( t\right) ,$ where $\eta $ is random with $E\eta \left(
t\right) =0,$ while $\bar{\xi}$ is the mean and $\sigma ^{2}$ is the
volatility. \bigskip Using Taylor series expansion of $f\left( t,\xi \left(
t\right) \right) $ around $\psi =\left( a,b\right) $ (and suppressing the
notational dependence on $t)$, we obtain

\begin{eqnarray}
f\left( t,\xi \right) &=&f\left( \psi \right) +f_{t}\left( \psi \right)
\left( t-a\right) +f_{\xi }\left( \psi \right) \left( \xi -b\right) +\frac{1%
}{2}f_{\xi \xi }\left( \psi \right) \left( \xi -b\right) ^{2}+\frac{1}{2}%
f_{tt}\left( \psi \right) \left( t-a\right) ^{2}+  \notag \\
&&f_{t\xi }\left( \psi \right) \left( t-a\right) \left( \xi -b\right)
+R\left( t,\xi \right) ,
\end{eqnarray}%
where $R\left( t,\xi \right) $ is the remainder and the subscripts denote
derivatives. Our intermediate objective is to minimize the absolute value of
the expected value of the remainder

\begin{equation*}
\underset{\bar{\xi}}{\min }\left| ER\left( t,\xi \right) \right| ,
\end{equation*}%
where

\begin{equation}
ER\left( t,\xi _{t}\right) =E\left[ 
\begin{array}{c}
f\left( t,\xi \left( t\right) \right) - \\ 
\left\{ 
\begin{array}{c}
f\left( \psi \right) +f_{t}\left( \psi \right) \left( t-a\right) +f_{\xi
}\left( \psi \right) \left( \xi -b\right) + \\ 
\frac{1}{2}f_{\xi \xi }\left( \psi \right) \left( \xi -b\right) ^{2}+ \\ 
\frac{1}{2}f_{tt}\left( \psi \right) \left( t-a\right) ^{2}+f_{t\xi }\left(
\psi \right) \left( t-a\right) \left( \xi -b\right)%
\end{array}%
\right\}%
\end{array}%
\right] .
\end{equation}%
The solution yields%
\begin{equation}
Ef_{\bar{\xi}}\left( t,\bar{\xi}^{\ast }+\sigma \eta \right) -f_{\bar{\xi}%
}\left( \psi \right) -f_{\bar{\xi}\bar{\xi}}\left( \psi \right) \left( \bar{%
\xi}^{\ast }-b\right) -f_{t\bar{\xi}}\left( \psi \right) \left( t-a\right)
=0.  \label{2}
\end{equation}%
Since the solution $\bar{\xi}^{\ast }$ depends on the values of $a$ and $b,$
the optimization is equivalent to choosing a specific value of $a=\hat{a}$
such that $\bar{\xi}^{\ast }\left( t\right) $ is equal to the actual
expected value of $\xi $ obtained from historical data or numerical methods.
Therefore the point of expansion $\hat{a}$ is not arbitrarily chosen
(however $b$ is arbitrarily chosen). Hence $\left( \ref{2}\right) $ can be
rewritten as%
\begin{equation*}
Ef_{\bar{\xi}}\left( t,\xi \right) =f_{\bar{\xi}}\left( \psi \right) +f_{%
\bar{\xi}\bar{\xi}}\left( \psi \right) \left( E\xi -b\right) +f_{t\bar{\xi}%
}\left( \psi \right) \left( t-\hat{a}\right) .
\end{equation*}%
Thus

\begin{eqnarray}
Ef\left( t,\xi \right) &=&f\left( \psi \right) +f_{t}\left( \psi \right)
\left( t-\hat{a}\right) +f_{\bar{\xi}}\left( \psi \right) \left( \bar{\xi}%
-b\right) +  \notag \\
&&\frac{1}{2}f_{\bar{\xi}\bar{\xi}}\left( \psi \right) E\left( \xi -b\right)
^{2}+  \notag \\
&&\frac{1}{2}f_{tt}\left( \psi \right) \left( t-\hat{a}\right) ^{2}+f_{t\bar{%
\xi}}\left( \psi \right) \left( t-\hat{a}\right) \left( \bar{\xi}-b\right) .
\label{1}
\end{eqnarray}%
Without loss of generality, we define $f\left( t,\xi _{t}\right) =Ef\left(
t,\xi \right) +\delta \left( t\right) \varpi \left( t\right) ,$ where $%
\varpi $ is random and $\delta ^{2}$ is the volatility of $f$. Adding $%
\delta \left( t\right) \varpi \left( t\right) $ to both sides of $\left( \ref%
{1}\right) ,$ we obtain 
\begin{eqnarray}
f\left( t,\xi \right) &=&f\left( \psi \right) +\delta \varpi +f_{t}\left(
\psi \right) \left( t-\hat{a}\right) +  \notag  \label{3} \\
&&f_{\bar{\xi}}\left( \psi \right) \left( \bar{\xi}-b\right) +\frac{1}{2}f_{%
\bar{\xi}\bar{\xi}}\left( \psi \right) E\left( \xi -b\right) ^{2}+  \notag \\
&&\frac{1}{2}f_{tt}\left( \psi \right) \left( t-\hat{a}\right) ^{2}+f_{t\bar{%
\xi}}\left( \psi \right) \left( t-\hat{a}\right) \left( \bar{\xi}-b\right) .
\label{3a}
\end{eqnarray}%
Differentiating $\left( \ref{3a}\right) $ with respect to $t$ and $\varpi ,$
respectively, we obtain%
\begin{equation}
f_{t}\left( t,\xi \right) =f_{t}\left( \psi \right) +f_{tt}\left( \psi
\right) \left( t-\hat{a}\right) +f_{t\bar{\xi}}\left( \psi \right) \left( 
\bar{\xi}\left( t\right) -b\right) +\delta _{t}\left( t\right) \varpi
_{t}\left( t\right) ,
\end{equation}%
\begin{equation}
f_{\xi }\left( t,\xi ^{\ast }\left( t\right) \right) =\delta \left( t\right)
\varpi _{\xi }\left( t\right) ,  \label{4}
\end{equation}%
and thus%
\begin{equation}
df\left( t,\xi \right) =\left[ 
\begin{array}{c}
f_{t}\left( \psi \right) +f_{tt}\left( \psi \right) \left( t-\hat{a}\right)
+f_{t\bar{\xi}}\left( \psi \right) \left( \bar{\xi}\left( t\right) -b\right)
+ \\ 
\delta _{t}\left( t\right) \varpi _{t}\left( t\right)%
\end{array}%
\right] dt+\delta \left( t\right) \varpi _{\xi }\left( t\right) d\xi \left(
t\right) .  \label{5}
\end{equation}%
If $t=0$ (time-independence), $\left( \ref{5}\right) $ is reduced to 
\begin{equation}
df\left( \xi \right) =\delta \varpi _{\xi }d\xi .  \label{6}
\end{equation}%
Therefore, virtually, the dynamics of any stochastic process are given by $%
\left( \ref{5}\right) $ or $\left( \ref{6}\right) .$ The extension to the
multiple-variable case is straightforward.

\textbf{3. Practical example (the investment model)}

We apply our new method to the standard investment model (a major model in
finance). Below is a brief description of the investment model (see, for
example, Focardi and Fabozzi (2004), among many others) The wealth process
is given by

\begin{equation}
X\left( T\right) =x+\int\limits_{s}^{T}\mu \pi \left( t\right)
ds+\int\limits_{s}^{T}\pi \left( t\right) \sigma dW\left( t\right) ;s\leq
t\leq T,
\end{equation}%
where $x$ is the initial wealth, $\pi $ is the risky portfolio (the value of
the risky asset), $W$ is a Brownian motion, $\mu $ is the risk premium (the
rate of return of the risky asset minus the risk-free rate), and $\sigma $
is the volatility of the risky asset. Hence,

\begin{equation}
dX\left( t\right) =\mu \left( t\right) \pi \left( t\right) ds+\pi \left(
t\right) \sigma \left( t\right) dW\left( t\right) .
\end{equation}%
The investor invests a fraction of his/her wealth in the risky asset and the
remainder in the risk free asset (such as a bank account). The investor's
objective is to maximize the expected utility of the terminal wealth with
respect to the risky portfolio

\begin{equation*}
V\left( s,x\right) =\underset{\pi }{Sup}E_{s}U\left( X\left( T\right)
\right) ,
\end{equation*}%
where $V\left( .\right) $ is the value function, $U\left( .\right) $ is a
differentiable and bounded utility function. Similar to previous literature,
we define $E_{s}U\left( X\left( t\right) \right) =J\left( t,X\left( t\right)
\right) .$

Now we apply our new approach. Using $\left( \ref{5}\right) $ and choosing $%
b=0,$ we obtain

\begin{equation}
dJ\left( t,X\right) =\left[ J_{t}\left( \psi \right) +J_{tt}\left( \psi
\right) \left( t-\hat{a}\right) +J_{tX}\left( \psi \right) EX+\delta
_{t}\left( t\right) dW\left( t\right) \right] dt+\varpi _{X}\left( t\right)
\delta \left( t\right) dX\left( t\right) .  \label{9}
\end{equation}%
Using Stein's lemma $\delta \left( t\right) =J_{XX}\pi ^{2}\left( t\right)
\sigma ^{2}\left( t\right) $ and $\varpi _{X}\left( t\right) =1/\pi \left(
t\right) \sigma \left( t\right) $, and since the partial derivatives in $%
\left( \ref{9}\right) $ are constant, we can rewrite $\left( \ref{9}\right) $
as

\begin{equation}
dJ\left( t,X\right) =\left( c_{_{1}}+c_{_{2}}\left( t-\hat{a}\right)
+c_{_{3}}EX\right) dt+c_{_{4}}\pi \left( t\right) \sigma \left( t\right)
dX\left( t\right) ,
\end{equation}%
where $c_{_{i}}$ is a constant.

\bigskip To solve for the optimal risky portfolio $\pi ^{\ast }$, using the
Feynman-Kac approach, we obtain

\begin{equation}
c_{_{1}}+c_{_{2}}\left( s-\hat{a}\right) +c_{_{3}}x+\underset{\pi }{Sup}%
\left\{ c_{_{3}}\mu \left( s\right) \pi \left( s\right) +c_{_{4}}\pi
^{2}\left( s\right) \sigma \left( s\right) \mu \left( s\right) \right\} =0.
\end{equation}%
Differentiating the above formula with respect $\pi \left( s\right) $ and
arranging yields

\begin{equation}
\pi ^{\ast }\left( s\right) =\frac{c}{2\sigma \left( s\right) },
\end{equation}%
where $c=-c_{_{3}}/c_{_{4}}.$ It is worth emphasizing that this solution is
far simpler than the solution based on Ito's calculus.

\bigskip \pagebreak

\end{document}